\documentclass[reprint,showpacs,twocolumn,a4paper]{revtex4}

\usepackage{graphics}%

\begin{document}

\title{\large Fabrication of quantum point contacts by engraving \\ GaAs/AlGaAs-heterostructures with a diamond tip}

\author{J. Regul}
\author{U. F. Keyser}
\author{M. Paesler}
\author{F. Hohls}
\author{U. Zeitler}
\author{R. J. Haug}
\affiliation{%
Institut f\"ur Festk\"orperphysik, Universit\"at Hannover, 30167
Hannover, Germany}
\author{A. Malav\'{e}}
\author{E.  Oesterschulze}
\affiliation{%
Institut f\"ur Technische Physik, Universit\"at
Kassel, 34132 Kassel, Germany}
\author{D. Reuter}
\author{A. D. Wieck}
\affiliation{%
Lehrstuhl f\"ur Angewandte Physik, Ruhr-Universit\"at
Bochum, 44780 Bochum, Germany}

\date{\today}

\begin{abstract}
We use the all-diamond tip of an atomic force microscope for the
direct engraving of high-quality quantum point contacts in
GaAs/AlGaAs-heterostructures. The processing time is shortened by
two orders of magnitude compared to standard silicon tips.
Together with a reduction of the line width to below 90~nm, the
depletion length of insulating lines is reduced by a factor of two
with the diamond probes. The such fabricated defect-free ballistic
constrictions show well-resolved conductance plateaus and the
0.7~anomaly in electronic transport measurements.
\end{abstract}


\maketitle
\newlength{\plotwidth}          
\setlength{\plotwidth}{7.5cm}


Over the last years the atomic force microscope (AFM) has been
used as a flexible nanolithographic tool for the direct patterning
of surfaces~\cite{afm_new}. It offers not only a convenient and
simple way to fabricate sub-micron devices but also permits in
situ control of relevant sample parameters during the lithography
process~\cite{schumach99}. A successful and straightforward method
is the mechanical manipulation of semiconductors surfaces by means
of an AFM-tip. The feasibility of this technique has been
demonstrated for various materials like GaSb~\cite{magno97},
InAs~\cite{cortesrosa98} and GaAs~\cite{schumach99,hyon00}.

Here we present the application of the engraving technique to
fabricate quantum point contact devices in
GaAs/AlGaAs-heterostructures. We show that new all-diamond tips
are ideally suitable for the manufacturing of defect-free
ballistic channels in two-dimensional electron gases. We consider
the importance of the AFM-tip material by comparing the device
properties of samples patterned by a silicon tip and by a diamond
tip. Because of its highest possible Mohs hardness of 10 diamond
represents the ideal tip material for the engraving. For the
patterning we use standard silicon tips~\cite{tips} and
all-diamond AFM-tips with force constants of more than 40~N/m. The
latter were grown by hot-filament chemical vapor deposition of
polycrystalline diamond onto a pre-patterned silicon substrate.
Details on the fabrication technique are given
in~\cite{oesterschulze}.
\begin{figure}
\begin{center}
 \includegraphics{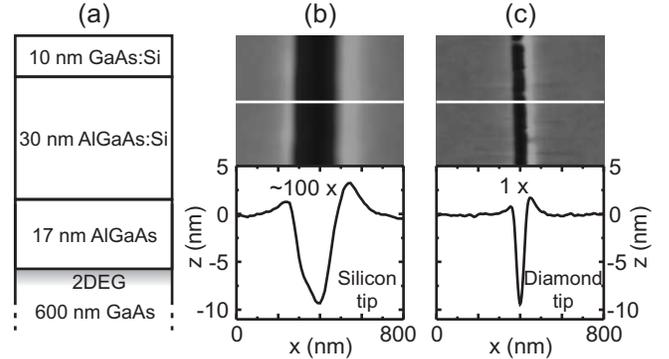}
\end{center}
\caption{(a) Layer sequence of the heterostructure. The doping
concentration of the two upper layers is $\sim
0.9\times10^{24}$~m$^{-3}$. (b)+(c) Results of the engraving with
(b) a silicon tip and (c) a diamond tip. Upper part: AFM
micrograph of the grooves. Lower part: depth profile along the
white lines. (b) After $\sim$100 scans with a silicon tip, (c)
after one scan with a diamond tip.} \label{abb1}
\end{figure}

The samples presented in this experiment are based on a modulation
doped GaAs/AlGaAs-heterostructure containing a two-dimensional
electron gas (2DEG) 57~nm below the sample surface with a sheet
density of 4.07$\times 10^{15}$~m$^{-2}$ and a mobility of
107~m$^2$/Vs, the layer sequence is shown in Fig.~\ref{abb1}(a).
We fabricated Hall bar geometries with standard photolithography,
wet-chemical etching and alloyed Au/Ge-contacts. Afterwards the
samples were bonded and mounted into the AFM for the controlled
engraving process. For the scribing the AFM-tip is repeatedly
scanned over the Hall bar with a scanning speed of 0.1~mm/s and a
contact force of several tens $\mu$N. Due to this high loading
force each scan removes some material of the cap layer which leads
to a stepwise depletion of the underlying 2DEG. During the whole
lithography procedure the sample resistance is monitored to
control the fabrication progress. The total depopulation of the
2DEG is marked by an abrupt raise of the sample resistance to more
than 3~M$\Omega$. For more details on our patterning procedure see
Ref.~\cite{schumach99}.

In Fig.~\ref{abb1}(b) we show an AFM-image of an engraved line
that was scribed with a Si-tip by applying 50~$\mu$N as loading
force and scanning the tip $\sim 100$ times over the surface. The
resulting line has a width of 250~nm. The depth $z \sim 9$~nm
suffices for this heterostructure for the total depletion of the
2DEG underneath the groove.

We achieve much narrower lines of 90~nm width and the same depth
$z \sim 9$~nm by using an all-diamond probe as shown in
Fig.~\ref{abb1}(c). The displayed groove was manufactured by
scanning the diamond tip once over the surface with a similar
contact force as for Si. Compared to the former results in
Fig.~1(b) the engraving process for e.g. a 100~$\mu$m line is
substantially reduced by nearly two orders of magnitude from
minutes to a few seconds. The reduction of the line width from
250~nm to 90~nm is mainly explained by the severe tip wear of the
Si-tip during the writing process. After the engraving we measured
the radius of the Si-tip and the diamond tip by scanning electron
microscopy. Whereas the Si-tip radius increased by a factor of 10
to more than 100~nm, images of the diamond tips yielded a radius
of below 50~nm before and after the fabrication. As expected the
tip wear for diamond is almost negligible. In fact, we used this
diamond tip for the fabrication of more than 40 devices without
any observation of tip degradation. In contrast, a silicon tip can
only be utilized once in most cases.
\begin{figure}
\begin{center}
 \includegraphics{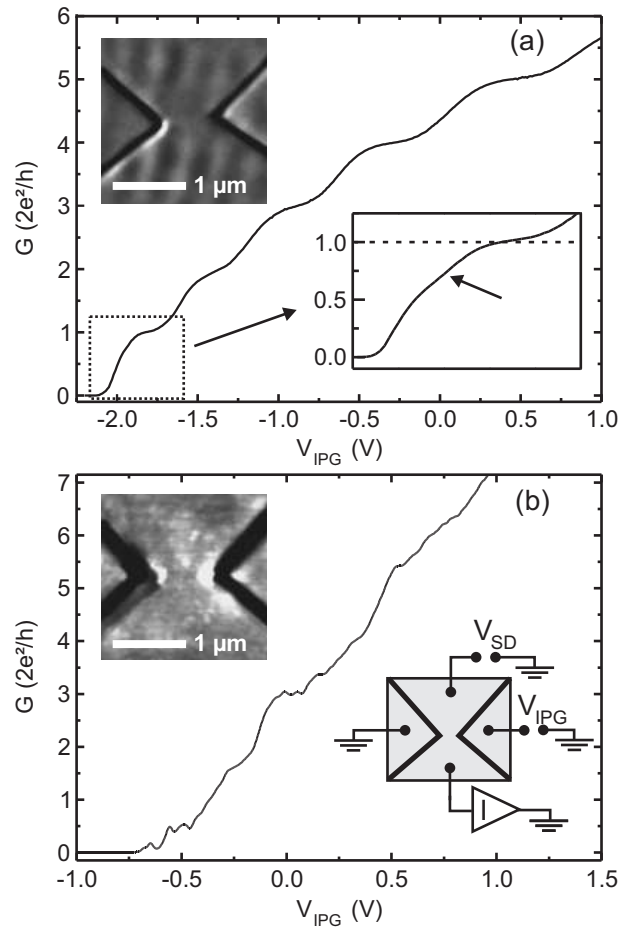}
\end{center}
\caption{Differential conductance $G(V_{IPG})=dI/dV_{SD}$ in units
of 2$e^{2}/h$ as a function of in-plane gate voltage: (a) Sample
patterned by a diamond tip. Left inset: AFM-image of a
constriction formed by a diamond tip. Right inset: Magnification
of the first conductance step. The arrow marks the 0.7 anomaly.
(b) Silicon-patterned sample. Left inset: AFM-image of a
Si-patterned constriction. The schematic on the right sketches the
measurement setup.} \label{abb2}
\end{figure}

To compare the electronic properties of the lines fabricated by
the different tips we defined two 1D channels by engraving
constrictions into the GaAs/AlGaAs-heterostructure. The regions
separated from the constriction by an insulating groove serve as
in-plane gates. The upper insets of Fig.~2(a) and Fig.~2(b) show
the constrictions engraved with the diamond tip and (b) the
Si-tip. Both constrictions were electrically characterized in a
pumped $^3$He-cryostat providing a base temperature of $T =
350$~mK.

In Fig.~2 the differential conductance $G=dI/dV_{SD}$ of the
diamond~(a) and the silicon-patterned sample~(b) is shown. For the
measurement we used a standard lock-in technique at an excitation
voltage of $V_{SD,ac} = 60$~$\mu$V at 13~Hz. The two conductance
curves presented in Fig.~2 were recorded by varying only a single
in-plane gate whereas the second gate was kept at a fixed
potential. A constant series resistance of the contacts and the
2DEG was subtracted. A schematic picture of the measurement setup
is shown in the lower inset of Fig.~2(b).

The curve in Fig.~2(a) corresponding to the diamond-patterned
sample shows flat quantized plateaus at multiple integers of
2$e^{2}/h$. This indicates the formation of a ballistic quantum
point contact~\cite{beenakker91} formed by an adiabatic potential
without any impurities. The appearance of the conductance plateaus
demonstrates that the grooves scribed with the diamond tip define
a smooth potential without significant fluctuations. In contrast,
the conductance of the silicon-patterned sample, shown in
Fig.~2(b), exhibits only a few poorly resolved conductance
plateaus.
\begin{figure}
\begin{center}
 \includegraphics{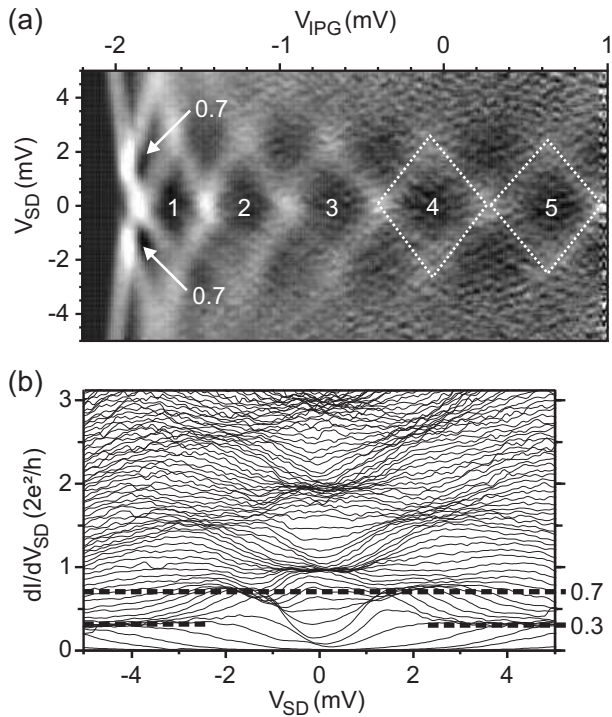}
\end{center}
\caption{(a) Grey-scale plot of the transconductance $dG/dV_{IPG}$
obtained from the diamond-patterned sample at a temperature of $T
= 350$~mK. The gate configuration is the same as in Fig.~2(b).
Dark regions correspond to low transconductance (plateaus) and
light regions reflect high transconductance (plateau transitions).
The numbers denote the occupied subbands. The location of the 0.7
plateaus is marked by the arrows. (b) Differential conductance as
a function of dc source-drain voltage taken at fixed gate
voltages. The additional plateaus at 0.7 are marked by a dashed
line. }\label{abb3}
\end{figure}

For further characterization of the diamond-patterned sample we
applied an additional dc source-drain bias voltage $V_{SD}$ that
allows us to determine the 1D-subband spacing~\cite{patel91}. The
transconductance $dG/dV_{IPG}$ derived numerically from these
measurements is plotted as a function of $V_{IPG}$ and $V_{SD}$ in
the grey scale plot in Fig.~3(a). The corresponding conductance is
given by the numbers inside the dark regions and the diamonds for
$G=4(2e^2/h)$ and $G=5(2e^2/h)$ are marked with dashed lines. The
crossing of adjacent zero-bias peaks $N$ and $N+1$ at finite bias
$eV_{SD}$ reveals the energy spacing $\Delta E_{N,N+1}= eV_{SD}$
ranging from $\Delta E_{2,3}=2.5(\pm 0.1)~\mathrm{meV}$ for the
second and third subbands to $\Delta E_{4,5}=2.3(\pm
0.1)~\mathrm{meV}$. Whereas the subband spacing in split gate
devices at higher subband indices drastically decreases we observe
only a slight reduction for our sample at $N>1$. This indicates
that the shape of confinement inside the constriction remains
nearly unaffected by the gate voltage and is only shifted up and
down. Assuming a harmonic confinement potential and a gate voltage
dependent potential barrier we deduce a value of $w~\sim$~160~nm
for the electronic width of the constriction at zero gate voltage
with $\sim$five occupied subbands. The depletion length for the
diamond tip then can be determined to $w_{depl}~\sim$~180~nm which
nearly is half the length of $w_{depl}~\sim$~330~nm extracted for
the silicon-patterned sample. The larger depletion length of the
silicon tip as well as the creation of significant potential
fluctuations are probably related to an enhanced formation of
surface defects caused by the increased number of scans.

By inspection of the first conductance step in the right inset of
Fig.~2(a) we observe an additional shoulder close to
0.7~(2$e^{2}/h$). In the grey scale plot in Fig.~3(a) this
shoulder leads to additional plateaus for finite bias voltages at
$G<2e^{2}/h$ marked with arrows. This can be seen more clearly in
Fig.~3(b), where we plotted $G=dI/dV_{SD}$ as a function of dc
source-drain voltage taken at fixed gate voltages. Whereas the
majority of the plateaus appear at multiples of $2e^2/h$, below
$2e^2/h$ extra plateaus appear at 0.3($2e^{2}/h$) and
0.7($2e^{2}/h$) marked by the horizontal dashed line. The
so-called 0.7 anomaly~\cite{thomas96} is an indicator for very
clean one-dimensional channels and is considered to be caused by
electron-electron interactions. The exact underlying mechanism of
this structure is still not clarified but it is an intrinsic
property of low-disorder quantum point contacts. Together with the
well-resolved plateaus the appearance of the 0.7 anomaly shows
that we scribed an adiabatic-like constriction free from
significant potential fluctuations with the diamond tip.

In conclusion, we fabricated quantum point contact devices by
engraving a constriction into a GaAs/AlGaAs-heterostructure with
the tip of an atomic force microscope. To study the influence of
the tip material we engraved devices using both a silicon tip and
a diamond tip. It turned out that a diamond tip is almost perfect
not only on the basis of a fast and simple processing but also in
forming proper potential profiles to observe ballistic electron
transport. The appearance of the 0.7~(2$e^{2}/h$) conductance
anomaly confirms the high-quality of diamond-engraved devices. We
deduced the depletion lengths induced by the different tips
yielding $w_{depl}~\sim$~180~nm for diamond-engraved samples which
is roughly two times smaller than typical depletion lengths
in silicon-patterned devices.\\
We thank P. Hullmann for his assistance with the scanning electron
microscope. This work was supported by the BMBF.

\end{document}